\documentstyle[12pt]{article}
\begin{document}
\newcommand{\etal}{{\em et al\/}}

\begin{center}
{\Large Deuteron -- $\alpha$ interaction by inversion of RGM $S$-matrix:\\
determination of spin-orbit potential for spin-1 projectile}

\vspace{20mm}

{\large R. S. Mackintosh~\footnote{To whom correspondence should be addressed.}
 and S.G. Cooper}%
%{\large S.G. Cooper and R.S. Mackintosh}%
\vskip1.0 cm%
{\large Physics Department, The Open University,\\ Milton Keynes
 MK7 6AA}

\vskip 0.5 cm%
r.mackintosh@open.ac.uk, s.g.cooper@open.ac.uk%
%s.g.cooper@open.ac.uk, r.mackintosh@open.ac.uk %

\end{center}

\vspace{20mm}

\noindent {\bf Abstract:} The iterative-perturbative (IP) procedure for $S$-matrix
to potential inversion is applied to spin-one projectiles for the restricted
case of vector spin-orbit interaction only. In order to evaluate this extension
of IP inversion we have inverted the  
multi-channel RGM $S_{lj}$ of Kanada
{\em et al\/} for deuterons scattering from $^4$He with deuteron distortion and
then compared the central components with those derived from RGM with spin
set to zero.  Attention is given to the question of how well the 
resulting potentials are established. Reliable spin-1 inversion is demonstrated. 
Results relating to inversion, to deuteron-nucleus interactions and to
RGM are presented and suggest the range of nuclear interaction information
which the procedure makes possible.\\[1 cm] 

\setlength{\parindent}{0.0 in}
\setlength{\parskip}{4 mm}



\vfill

\hfill\today
\pagebreak

\section{Introduction}
For some time it has been straightforward to apply $S$-matrix to potential
inversion to spin half projectiles, i.e., symbolically, perform
the transformation $S_{lj} \rightarrow V(r) + {\bf l\cdot s} V_{\rm so}(r)$.
The same transformation for spin one particles is much more problematic, not least
for the number of possible tensor forces, some of which couple certain
projectile partial wave channels. Nevertheless, under certain circumstances,
inversion for spin one projectiles should be possible, and one purpose of
this paper is to
present some specific cases in which previously published $S_{lj}$ from RGM
calculations of deuteron-$^4$He scattering are inverted. 
To assist in the  evaluation of the resulting potentials, we also present
potentials inverted from $S_{l}$ from models which are similar except
that the deuteron is treated as spinless.  

Accordingly, we apply $S$-matrix-to-potential inversion
to spin one projectiles for the restricted case where it is reasonable 
to assume that there is no
tensor interaction. The $S_{lj}$  are from
multi-channel resonating group model (MCRGM) calculations. 
The algorithm which we employ  
is a natural extension of the IP method which has been
successfully applied to  spin-half $S_{lj}$, see 
Refs.\cite{invprob,prc54} and references therein.
In Section 2 we explain the particular features of the source
of $S_{lj}$ which make an inversion involving only a vector 
spin-orbit interaction appropriate. In this paper we invert only
$S_{lj}$ which have been published elsewhere and we discuss only such
aspects of the RGM calculations which are specifically relevant.
An exhaustive account of what can be learned
by applying inversion to RGM $S$-matrix requires RGM calculations
of a complexity well beyond those leading to the $S_{lj}$ which
we invert. Nevertheless, the results presented here indicate
what might be discovered  from systematic studies linking inversion
and RGM and similar theories. 
In another publication~\cite{shirley}, deuteron-$^3$He potentials based
on RGM are presented.

Inversion is not an end in itself and one of our goals
is to establish generic properties of nucleus-nucleus potentials. It can also
 help to evaluate RGM and similar calculations, which, for all their 
complexity, are in general not yet capable of giving precise fits
to experimental data. A particular strength of the RGM, of course,
is that it contains an exact treatment of exchange terms and centre of mass motion. 
It is therefore not surprising that many of our findings relate to
non-local effects and parity dependence, both of which have their origin in
the exchange terms. In particular, we present evidence, based on an analysis
of the imaginary components, for a form of non-locality very different
from that which is dominant in nucleon scattering. This is in addition to the parity 
dependence which, as in other cases studied by inversion, turns out to be rather 
different from what has been assumed in phenomenological studies.
 The present inversions not only yield
certain properties of deuteron-nucleus potentials, but suggest
more general properties of nucleus-nucleus interactions.

Section 2 presents relevant details of the MCRGM calculations. Section 3 describes the
generalisation of IP inversion to spin-1. Section 4 presents certain results
for spin-0 deuteron inversions principally to help  evaluate  the 
spin-one inversions described in Section 5. Section 6 presents, for comparison
with Section 5, potentials derived from earlier no-distortion spin 1 RGM. Section 7
summarises our conclusions 
concerning 
the inversion procedure itself and also presents certain findings about
d-$\alpha$ potentials and discusses future work.

\section{The Multi-channel RGM of Kanada {\em et al}}
Kanada {\em et al\/}~\cite{kkst} (hereafter KKST) 
calculated d + $\alpha$ elastic scattering at several energies
using  multi-channel RGM (MCRGM) in order to study specific distortion effects. 
Spin-orbit but not tensor nucleon-nucleon forces were included
and there was no coupling between $l= J-1$ and $l= J+1$ channels.
For this reason, the calculated complex phase shifts were 
enumerated according to $l$ and $J= l, l\pm 1$. The dynamic distortion
of the deuteron was simulated by coupling to bound excited quasi-states
of the deuteron. This coupling generates absorption, 
but insufficient absorption for the data to be fitted, $^3$He $+$ t 
channels being omitted for example. No D-state components were included in the 
deuteron and the coupling was exclusively to S-wave virtual states of
the deuteron. In lowest order folding 
model, see Ref.~\cite{ka} and references therein, it is the D-state
which is responsible for the $T_R$~\cite{grsnp21} tensor term.\footnote{The 
full classification, based on symmetry principles, of the possible
tensor forces was given by Satchler~\cite{grsnp21}, who 
 defined  the three terms $T_R$, $T_P$ and
$T_L$. See also Robson~\cite{bar}.}

The $T_P$ tensor term is due to Pauli blocking~\cite{andy}, and  this
might be expected to be represented in fully antisymmetrized
RGM. However since the $T_P$ term depends on the 
suppression of effects due to the
tensor component of the nucleon-nucleon interaction, see Ref.\cite{andy},
the KKST  $S$-matrix may not contain significant effects requiring
 a $T_P$ interaction. The diagonal $T_L$ interaction 
is expected to be small on general grounds~\cite{stamp}.
 
The above features suggest that it is reasonable to represent the
resulting $S_{lj}$ with no more than a vector spin-orbit interaction
added to the central potential.  

KKST included an additional local and $l$-independent phenomenological
imaginary potential in order to simulate the absorption from the
excluded channels. Adjustment of the parameters of
this imaginary potential led to acceptable fits to the angular distributions,
but neither the tensor nor vector analysing powers were better than 
qualitative, the vector analysing powers being, unsurprisingly,  more 
acceptable. The complex phase shifts were tabulated for CM energies
of 8.0, 14.0, 19.6 and 37.33 MeV. In this paper we present and discuss
potentials for the 19.6 and 37.33 MeV cases.

\section{Inversion for spin-one projectiles}
Inversion for spin-zero projectiles, involving the $S_l \rightarrow V(r)$
transformation, is now straightforward, and a variety of methods
are available, see Refs.~\cite{chadsab,badhon}. 
Here we use  the iterative perturbative, IP, method~\cite{invprob}; see
Ref.~\cite{prc54} for recent developments.

The IP method allows inversion
for spin-half particles, i.e.\ $S_{lj} \rightarrow V(r) + {\bf l \cdot
\sigma} V_{\rm so}(r)$. Although  the results are most
 naturally presented in terms of a central and spin-orbit
component, one can think of this as equivalent to
inverting the two sets of amplitudes for $j=l\pm 1$, leading to two 
corresponding potential components. 
The spin-one situation is  much more complicated since there are
three values of $l$ for each $j$ and, in general, coupling between
$l=j-1$ and $l=j+1$ partial waves. The inversion of the full
non-diagonal $S$-matrix to yield the general sum of vector and tensor 
components~\cite{grsnp21} is very challenging. Indeed, inversion
from the  incomplete and imprecise non-diagonal $S$-matrix as
generally determined empirically involves major ambiguity problems.

The calculations of KKST 
present the opportunity and motivation for
a more restricted form of spin-one inversion. A straightforward extension of the
IP method will be applied to the KKST $S_{lj}$ yielding
a potential of the form $V(r) + {\bf l \cdot
S} V_{\rm so}(r)$. Three sets of amplitudes 
determine two potential components, whereas for spin-half
two sets of amplitudes determine only two potential components.
It is not {\em a priori\/} obvious that this will be possible, at least
yielding reasonably smooth potentials. Since $T_R$, $T_P$ and $T_L$ all
have diagonal matrix elements, an arbitrary
set of   $S_{lj}$ might  contain  effects
which require representation by the diagonal parts of those components.
In our restricted model, any  effects which would 
be attributable in a full 
model to the diagonal tensor components will tend to induce
`waviness' in the potentials.

Since we must allow potentials fitting RGM $S_{lj}$ to be parity 
dependent, there are twice as many components to be 
determined. The notation we have used
previously~\cite{prc54} is that each real or imaginary, central
or spin-orbit term has two components  referred to
as $V_1$ and $V_2$ components, defined as $V(r) = V_1(r) + (-1)^l V_2(r)$.

The combined effects of parity dependence, the limits on the numerical precision of 
the KKST
$S_{lj}$ and the uncertainty as to whether an ${\bf l \cdot S}$ term is
sufficient, are such that ambiguity problems could well be serious. An advantage of the
IP procedure, compared to methods which simply take $S_{lj}$,
`turn a formal crank' and get $V(r)$, is that there is control
over the ambiguities.
It is actually quite important that {\em exact\/} inversion is avoided
in cases where $S_{lj}$ contain significant `noise' due to numerical
limitations ($S_{lj}$ from theory) or fitting errors (empirical $S_{lj}$).
 Over-fitting is avoided by controlling the dimensionality of the inversion
basis and adjusting the  SVD singularity parameter (see Ref.\cite{prc54}
and references therein). Although 
the criterion for a `smooth'  potential is inevitably somewhat subjective,
 consistent results can be achieved. All the potentials presented
below correspond to very low values of the convergence parameter
$\sigma$ defined~\cite{invprob} as \begin{equation}
  \sigma^2 = \Sigma_{lj} |S_{lj}^t -S_{lj}^i|^2 \end{equation}
where $S_{lj}^t$ is the `target' $S$-matrix for which we seek the 
potential, and $S_{lj}^i$ is the $S$-matrix of the inversion potential. Unless we
indicate otherwise, the potentials we present will be such that
$S^t$ and $S^i$ would be indistinguishable on a graph, or very nearly so.
Unfortunately, there can be a surprising range of potentials associated
even with such low $\sigma$. This is true even with the extreme low $\sigma$ 
(which we carefully avoid) associated with over-fitting numerical noise.
We therefore study the uniqueness of inverted potentials by varying the 
`starting potential'  and the basis in which the final potential is expanded.
Further substantiation comes from comparison  with d-$\alpha$ potentials derived 
from RGM calculations in which spin degrees of freedom are omitted.
                               
\section{Conclusions from spin-zero deuteron models} Uncertainties connected
with the omission of tensor forces suggest  that, in attempting
to establish spin-one deuteron-nucleus potentials, we should compare
the spin-independent parts of the potentials with  potentials found in studies 
where the complication of spin is avoided. We therefore describe
some results from calculations 
involving  $S$-matrices from RGM calculations in which the deuteron
is treated as spinless. We find, {\em inter alia\/},
(i) that we must not omit  parity dependent components of the potential,
and, (ii) that the exchange kernel will have an effect
on the imaginary potential unlike that for nucleon scattering.
\subsection{Real potential, spin-zero case}
We inverted phase shifts~\cite{yct-priv} for 20, 40, 60 and 80 MeV CM
d + Alpha RGM calculations in which
Coulomb and spin-orbit terms were omitted. There was no channel
coupling and no phenomenological representation of absorption, so 
the resulting potentials were real.
For each energy, we had  real phase shifts 
for $0 \le l \le 8$ so that at the two highest energies, the phase shifts
for the highest available partial waves were far from zero. 
For this reason, there is no meaning
attributable to $V(r)$ for $r>R(E)$, where $R(E)$ is the upper 
radius over which the inversion basis was defined at energy $E$;
$R(20) =R(40) = 8$ fm, $R(60) =7$ fm and $R(80)=6$ fm. These values
were found empirically by  requiring that smooth potentials
should exist.

The relevant finding is that when all the exchange 
terms, explicitly including the `2-exchange'~\cite{npa320} terms,
are included in the RGM calculation, the local equivalent potential
is substantially parity dependent. Although the parity dependent component
falls with energy, it is  appreciable at 80 MeV CM. We therefore
expect such terms in the lower energy spin-dependent cases  discussed below.
The parity dependent potential $V_2$ does not follow the form of the parity
independent term, but is surface peaked. It is such that the
the {\em even} potential is deeper,
notably for 2---5 fm. For 20 MeV deuterons, $V_2$ is attractive so that
at 4 fm, where $V_1\sim5$ MeV, the ratio of even to odd potential
is about 2:1. The effect falls with energy
but the qualitative pattern is retained.
At 60 MeV the $V_2$ term is largest around 2.5 fm at which radius
 the even parity potential is some 2.5 MeV deeper
than the odd parity potential. We conclude 
that when spin is included, we must allow 
 all components of the potential to be parity dependent.

The same $S$-matrix elements can also be subjected to energy dependent 
inversion~\cite{prc54}
in which the potential is expressed as a product of energy function and
a fixed radial form. The present data are well represented by\begin{equation}
 (1- 0.00237 E) V_1(r) + (-1)^l(1 - 0.00919 E) V_2(r) \label{eq20-80}
\end{equation} where
$V_2(r)$  is a surface peaked attractive term with a maximum magnitude
at $r=3$ fm where its value is $-2.2$ MeV. The qualitative features of
the potential are just those found  with fixed energy inversion 
apart from the change in radial form with energy.
Equation~\ref{eq20-80} implies 
 that the parity dependent term becomes zero
around 109 MeV, so we must presume parity dependence
for both energies in the spin-1 inversions.  

\subsection{Complex deuteron potential at 55 MeV}
Thompson and Tang~\cite{tt-prc8} calculated d + $\alpha$ elastic scattering 
at 55 MeV CM
using no-distortion RGM with a local $l$-independent phenomenological 
imaginary potential to simulate absorption. They adjusted the parameters of
the imaginary potential to achieve a reasonable fit to the elastic 
scattering differential cross section. No spin effects were included.

The tabulated $S_l$ of Thompson and Tang were inverted  and the salient 
features of the resulting potentials were as follows:\begin{enumerate}
\item Smooth potentials closely reproducing $S_l$ could be found if and 
only if the potential had a substantial parity dependent ($V_2$) component.
The properties  of the parity dependent potential were well-determined.
\item In the outer radial region, $r\ge 2.5$ fm,
 the imaginary $V_1$ potential follows quite closely the local imaginary
potential (`bare' potential) incorporated in the RGM.  However, towards the nuclear 
centre, particularly for $r\le 1.5$ fm, it is much
{\em deeper\/} than the bare potential. 
\item The real $V_2$ term is purely attractive, has a maximum magnitude
of 2.5 MeV at about 1.8 fm and is near zero at the origin. It is therefore
in essential respects the same as in the real potential case in Section 4.1.
\item The imaginary $V_2$ term is  repulsive, peaking near 1.5 fm with a 
maximum of about 1 MeV. 
\item The volume integral of the real $V_1$ component is
well determined  by inversion and is reasonable at
470 MeV fm$^3$ per projectile nucleon at a laboratory energy of 82.5 MeV  
(41.25 MeV per projectile nucleon.)
\end{enumerate}

If there were no exchange terms and the consequent non-locality, the inversion 
should yield an imaginary component
precisely equal to the added phenomenological imaginary potential. 
But this is not found, the imaginary potential
being substantially {\em enhanced\/} at the nuclear centre. 
Remarkably, the effect of Perey-type non-locality is the
reverse~\cite{mc}: this is an `anti-Perey effect'. 
This is a firm conclusion; for example, the form of the imaginary $V_1$ component 
is essentially the same even when the parity dependent 
($V_2$) components are excluded. (Such inversions
lead to much larger values of 
$\sigma$ but yield $V_1$ components with similar properties.)  
It seems that the form of non-locality responsible for this effect is
independent of the exchange processes which lead to $V_2$ (parity 
dependence) terms. The `Perey-like' effects which arise in the presence
of single nucleon `knock-on' exchange are discussed, for example
in Ref~\cite{mc}. In the presence of a Perey-Buck non-local potential~\cite{pb}
which simulates very well the knock-on exchange seen by single nucleons~\cite{mc},
one finds that inversion leads to a uniformly {\em reduced\/} imaginary potential,
very satisfyingly consistent with what one would expect on the basis of a conventional
Perey~\cite{perey} effect. The existence of the reverse effect in RGM calculations
of the scattering of $A\ge2$ projectiles is therefore  striking evidence for the
very different action of other~\cite{npa320} exchange terms in the kernel.

The character of the $V_2$ term is
consistent with the  results of Section 4.1: $V_2$ is essentially 
 surface peaked  corresponding to additional attraction for the 
even parity partial
waves. Its maximum value, about 2.5 MeV, is consistent with the 60 MeV 
case of the last section, but the maximum is at  about 1.7 fm, whereas
the radius of the maximum was about 2.5 fm for the 60 MeV case.

\section{Inversion of KKST complex phase shifts}
We present results for the 37.33 MeV case before those for the 19.6 MeV 
case. At the lower energy there is  considerably
less information to determine the potential and it is appropriate
to exploit information from the higher energy case. 
At 37.33 MeV, $S_{lj}$ were calculated for $l\le 10$ but at 19.6 MeV,
only for $l \le 8$.

The behaviour of $|S|$ in Figure 6 of KKST suggests a much greater imaginary 
potential at the higher energy, a natural consequence of
the larger number of open channels.  It also suggests a substantial
imaginary spin-orbit component at both energies since $|S_{lj}|$  
 is much smaller for $j= l+1$ than for
$j=l-1$. The difference is too great to be a second order 
consequence of the real spin-orbit potential.
 
\subsection{Deuteron-$\alpha$ interaction at 37.33 MeV CM}
For each $l >0$, there are 3 complex 
phase shifts. Hence, with $l\le10$, the IP procedure can handle up to
62 real basis functions before the linear system becomes under-determined. 
This means that of the four complex components, central and spin-orbit terms,
each with parity independent and parity dependent components, three can have 
 bases of dimensionality 8 and one of 7. 
Since there are phase shifts for three $j$ values for each $l$, we might not
be surprised to find that spin-orbit components required bases of the higher
dimensionality (and so it turned out). For most of the calculations
we used a harmonic oscillator basis~\cite{kuku}
 with $R_0 =1.48$ and we assumed the potential was zero beyond
$R_{\rm max}= 7.5$ fm. Even with 31 basis functions 
for both real and imaginary components we could not achieve the `perfect
fit' low $\sigma$
values achieved for the zero spin 55 MeV case, so the final potentials
are less unique.  

A potential, `pot3', was found for which the larger components 
were free of waviness and for
which $\sigma = 0.00173$. There was  no discernible difference between
$S^t$ and $S^i$ when plotted. Alternative potentials were
found when using different starting potentials or different bases, but potentials for
which $\sigma$ was lower looked like pot3
but with superimposed waviness due to over-fitting. 
Figure 1 presents the complex central and spin-orbit $V_1$ components together
with those for  19.6 MeV discussed below, and Figure 2 presents the
corresponding $V_2$ (parity dependent) components. Table 1 presents volume integrals
and RMS radii for the real central $V_1$ parts of
 `pot3' and also of a somewhat more oscillatory potential, `pot1'. Potential pot1 
gave a poorer fit to $S_{lj}$ than pot3 as can be seen from $\sigma$.
Most components of pot1 exhibited typical over-fitting 
oscillations about the pot3 forms, yet the volume integrals and RMS radii 
were almost the same.\footnote{The somewhat paradoxical situation where pot1 
shows some waviness indicative of over-fitting yet gives worse
$\sigma$ than pot3 is due to the fact that certain small components were different 
in form, and
the oscillations in the larger components were induced by the iterative procedure 
trying to lower $\sigma$ with apparently incorrect solutions for the 
smaller components.} The volume
integrals of the larger components of pot3 are consistent with global optical
potentials. We note the following features of pot3:    
\begin{enumerate}
\item The imaginary $V_1$ potential follows the phenomenological imaginary
potential (not shown in Figure 1)
 quite closely for $r\ge 2.5$ fm. However, towards the nuclear 
centre, particularly for $r\le 1.5$ fm, it becomes very much
{\em deeper\/} than the bare potential to a much greater degree than for the 55 MeV case.
This is not solely an effect of the non-locality as it was for 
55 MeV
since the multi-channel RGM itself does contribute via deuteron 
breakup. Remarkably, the open channels make virtually
no contribution to the imaginary $V_1$ for $r \ge 2.5 $ fm, something also
found~\cite{shirley} in the d $+ ^3$He case.
The absorptive potential induced by these coupled channels has a very deep
minimum, about 25 MeV deep, at the nuclear centre. 
The non-zero slope of $V_1$ at $r=0$ indicates that it is a local 
representation of a non-local or $l$-dependent potential.
(In this  context, by $l$-dependence we mean variation across $l$-values
of particular parity, certainly possible for potentials 
generated by inelastic processes.)
\item The real $V_2$ term is mostly attractive with a local a maximum magnitude
of 2 MeV near 2.7 fm. It approaches  zero at 1.8 fm and is quite large near
 the origin. This is very similar to the 55 MeV $V_2$ term and consistent with
the cases mentioned in Section 4.1.
\item The imaginary $V_2$ term is absorptive.
It is not  much like the corresponding 55 MeV term, but this might 
be due to the fact that here the multi-channel RGM itself 
gives rise to absorption.
The $V_2$ term is almost a third of the magnitude of 
$V_1$ term near 2.5 fm, the  radius at which the bare
potential peaks and where, see point 1, $V_1$ hardly differs from
the bare potential. 
\item The volume integral per target nucleon  
of the real $V_1$ component, $J_{\rm R}$, is well determined at about
480 MeV fm$^3$ per projectile nucleon. This is phenomenologically
reasonable for a laboratory energy of 56 MeV  
(28 MeV per projectile nucleon.) 
\item The imaginary spin-orbit terms are quite small but not zero. 
The parity dependent terms are generally small except near the origin.
\item Although the real spin-orbit term is smooth, it is not known why it should
be peaked at the nuclear centre rather than being of Thomas form.  
Possibly  it is simulating a $T_L$ term. 
\end{enumerate}

How well-established are these properties? In order
to study the uniqueness of the potential, we performed
inversions with an alternative  inversion basis (Gaussians)
and with alternative starting potentials. In Figures 3 and 4, we compare
pot3 with potentials `potG1',
the lowest-$\sigma$ solution found using a Gaussian
basis that was not conspicuously
wavy. The characteristics of the central $V_1$ components of
potential potG1 are included in Table 1, with $J_{\rm R}$, $r$-RMS and $J_{\rm I}$ 
being essentially the same as for pot3.
The larger-magnitude components visible in Figures 3 and
4 show that `potG1' is essentially pot3 with superimposed waviness, so that 
in view of the substantially larger  $\sigma$, the comment
in the  footnote 3 applies.
Since we find quite
small imaginary spin-orbit components, we inverted with  imaginary 
spin-orbit terms held zero, and found
that such terms were essential, even though they are small and less well-determined 
than other terms.  We conclude that pot3 is a reliable
representation of the local potential corresponding to the 37.33 MeV KKST
$S_{lj}$.

In Table 1 we include for comparison  the phenomenological 
optical
potential of Hintenberger {\em et al\/}~\cite{hinten} for deuterons scattering
from $^4$He at 52 MeV laboratory energy.

\begin{table}[h] \caption{Properties of central potentials 
($V_1$ terms) for 37.33 MeV  and 19.6 MeV deuterons, KKST $S$-matrix..
OM refers to the optical model of Hintenberger{\em et al\/}}\begin{center}
\begin{tabular}{|cccccc|}\hline 
 Solution& $\sigma$ &$J_{\rm R}$& $r$-RMS (real)&$J_{\rm I}$& $r$-RMS (imag)\\ 
\hline
\multicolumn{6}{c}{37.33 MeV}\\ \hline
pot1 & 0.00320 & 478.15 & 2.960& 136.22& 3.487\\
pot3 & 0.00173 & 479.86 & 2.962& 136.85 & 3.463 \\ 
potG1 & 0.00486 & 483.33 & 2.961 & 137.56 & 3.429 \\ 
OM & --- & 539.48 & 2.68 & 170.66 & 3.211 \\ \hline
\multicolumn{6}{c}{19.6 MeV}\\ \hline
pot-a & $0.903 \times 10^{-3}$ & 499.72 & 2.905& 101.75& 3.448\\ \hline

\end{tabular}\end{center}
\end{table}  

\subsection{Deuteron-$\alpha$ interaction a 19.6 MeV CM}
As noted above, there are fewer $S_{lj}$  to fix the same number of potential
components. The  37.33 MeV potentials were
used as starting potentials.
Figures 1 and 2 show  components of `pot-a' which reproduces the 
19.6 MeV KKST $S_{lj}$ with
 $\sigma = 0.000903$. Characteristics of pot-a are given in Table 1. 
We find:
\begin{enumerate}
\item The real, central $V_1$ component is smooth and
substantially deeper at the nuclear centre than for 37.33 MeV. This reasonable depth
dependence is largely due to exchange  since the
nucleon-nucleon interaction is fixed. The volume integral is correspondingly
larger than for 37.33 MeV.
\item A remarkable  difference is the complete absence of 
one of the most striking features in the 37.33 MeV case: the deep central
dip in the imaginary central $V_1$ component. There is a corresponding
30\% decrease in the volume integral. We attempted unsuccessfully to find
potentials which did not have this property.
\item The real central $V_2$ term is more attractive than for 37.33 MeV except
for $r<1$ fm. This is consistent with the trend found in Section 4.1 
that $V_2$ is attractive and increases
with decreasing energy.
\item All the other imaginary components are substantially different at
the two energies.
\item The real spin-orbit components, both $V_1$ and $V_2$, are remarkably
similar at the two energies.
\end{enumerate} 

Concerning point 2: The difference between the 19.6 MeV and 37.33 MeV potentials 
is not surprising in view of the considerable difference,
clearly evident in KKST, between $|S_{lj}|$ for 19.6 and
37.33 MeV, including the large spread of values for different $j$ for given $l$.
There is much to be learned as to how the imaginary potential 
develops with energy as  channels open and this question bears upon
our general  understanding of how inelastic processes give rise to absorption. 

Just as for 37.33 MeV, various tests were performed to 
verify that the potential is reliably determined. We established, too,
that an imaginary spin-orbit potential is required, though the relative
accuracy of this small term is less than for the other components.

\section{$S$-matrix of Lemere, Tang and Thompson}
The earlier study of d + $\alpha$ scattering by Lemere {\em et 
al\/}~\cite{ltt}
omits all deuteron excitation but does include spin-orbit terms in the NN interaction
and, moreover, covers a considerable energy range. Although a somewhat different NN
interaction was used, it presents an opportunity to disentangle deuteron distortion
and spin-orbit effects. We determined 
potentials corresponding to their RGM $S_{lj}$ for CM energies of
16.58 MeV, 24.87 MeV and 53.9 MeV (i.e.\ up to 80.85 MeV
laboratory energy.) Absorption was represented solely by
a local imaginary potential\footnote{For other details see
Lemere {\em et al\/}; we are grateful to Professor Tang for bringing this 
to our notice and suggesting the present calculations.}. For 53.9 MeV,
only $J$-averaged phases were provided for the highest $l$-values, i.e.\
$l=9$, 10, and  11. 

As with other spin-1 inversions, ambiguities arise, particularly
if the restriction to reasonably smooth potentials is abandoned in pursuit of
very low $\sigma$. Again, this is
probably due to the inadequacy of the ${\bf L \cdot S}$ form fully to represent
the $lj$ dependence implicit in the $S_{lj}$. 
In order to establish reliable potentials, we applied
alternative basis sets etc.\ to the 24.87 MeV  $S_{lj}$. 
The optimum solution at 24.87 MeV served as a starting potential 
at 16.58 and 53.9 MeV. We considered it important to achieve
a uniform set of potentials for the three energies, with at least the largest
component (the real, central $V_1$ term) changing in a regular fashion over the
energy range, and only the smallest components (such as the imaginary $V_2$ spin-orbit
term) varying greatly from case to case. This makes it possible
to attribute physical meaning to the energy dependence of the large components.  

The optimum potentials are presented in Figures 5 and 6.
Volume integrals and rms radii of the real central $V_1$ term are presented
in Table~\ref{tablemere} with corresponding values of $\sigma$. Two
alternative solutions for 16.58 MeV are presented, giving
some idea of how well determined the potential is. The larger  $\sigma$ at
 53.9 MeV  is significant;  much  lower values corresponding
to very wavy potentials were rejected. In part, the larger $\sigma$
reflects the larger number of terms in the sum in Equation 1. Possibly, 
 the inadequacy of the simple spin-orbit form is 
greatest at the highest energy,  evidence for this being
that, although the value of $\sigma$ given for 53.9 MeV in Table~\ref{tablemere}
corresponds to a visually perfect fit to $S_{lj}$, the discrepancy in the observables
is noticeable, particularly in the tensor analysing powers. We may also
be seeing a consequence of the $j$-averaged $S_{lj}$ for $l \ge 9$. 
At 16.58 and 24.87 MeV, the inversion potentials give phase shifts yielding
observables which are graphically indistinguishable from
those calculated directly from the RGM  $S_{lj}$. 

Despite the higher $\sigma$ at 53.9 MeV, regular behaviour is apparent.
 However, the nature of the energy dependence is unexpected.  We see
from Figure 5 that the depth of the real, central $V_1$ component at
the nuclear centre falls steadily with energy as did the same term in Figure 1. 
But there is no corresponding
fall in the volume integral $J_{\rm R}$ in Table~\ref{tablemere}. 
Figure 5 shows this component  systematically 
{\em increasing\/} with energy in the surface region, a phenomenon reflected 
in the steady increase in rms radius, Table~\ref{tablemere}. 
The physical reason for this may well be an 
effect of two-nucleon exchange processes, an effect which
seems to have been compensated for 
in the calculations of the previous section by increasing surface repulsion due to
deuteron distortion. 
The fall with energy of the  potential at the nuclear centre 
is what would be expected from  
single nucleon knock-on exchange  (in the absence of channel coupling,
the energy dependence of the  real potential is entirely due to exchange.) 
Although the NN interaction of Lemere {\em et al\/} differs somewhat from that of
KKST, the difference  is unlikely to account for the different surface
behaviour of the real potential. 

A further clue to the nature of the exchange processes lies in 
the imaginary central $V_1$ potentials. In the RGM calculations of Lemere
{\em et al\/}, the absorption is entirely due to an added local imaginary potential.
If the non-local kernels were omitted from the RGM, the inversion would, of course,
reproduce the local kernel and the added local imaginary potential 
terms (with zero $V_2$ terms). The very different imaginary $V_1$ potentials displayed
in  Figure 5 are a consequence of the fact that the local imaginary potential is 
`seen' by a wave-function which is itself largely determined  by a non-local 
effective potential.Such behaviour was found already in section 4.2 where we 
discussed `anti-Perey' behaviour in the 55 MeV spinless case.
The behaviour shown in Figure 5 is also `anti-Perey' at the nuclear
centre to an extent which increases with energy.  Comparing the imaginary
central $V_1$ potentials with the  imaginary potentials included in Ref~\cite{ltt}
reveals a Perey-like reduction in the surface region, but an enhancement
at the nuclear centre. This occurs at each energy, and varies systematically,
with the smallest surface reduction and central enhancement at the lower
energies, and a large (Perey-like) surface reduction but small enhancement for
$r \le 1.5$ fm at 53.9 MeV.  Referring back to Figure 1, we see that
there is a similar strong anti-Perey effect at 37.3 MeV but not at 19.6 MeV.
Clearly, apart from generating $V_2$ potentials, those exchange terms
which are not of the one-nucleon (knock on) exchange type also have a substantial
effect on the interaction of deuterons with nuclei.

One of the most surprising features which was found for KKST potentials at both 
energies,
the deep real $V_1$ spin orbit potential, occurs consistently in Figure 5.

In summary, certain  striking features of the KKST potentials recur for
the case without deuteron distortion, but there certain differences
which suggest that a systematic study involving extended
RGM calculations will yield much interesting insight into
the combined influence of antisymmetrization and channel coupling
to inter-nuclear interactions.

\begin{table} \caption{\label{tablemere} Characteristics of real central $V_1$ 
potential reproducing $S_{lj}$ of Lemere {\em et al\/.}}
\begin{center}
\begin{tabular}{|cccc|}\hline 
 CM energy & $\sigma$& $J_{\rm R}$& $r$-RMS (real)\\ \hline
16.58 & 0.00314 & 543 & 3.079 \\ 
16.58 & 0.00136 & 552 & 3.070\\
24.87 & 0.00767 & 552 & 3.111 \\
53.9 & 0.0173 & 557 & 3.137\\ \hline
\end{tabular}
\end{center}
\end{table}

\section{Discussion, conclusions and future work}

{\bf Conclusions relating to inversion.}\,\,
The inherent flexibility of the IP algorithm has allowed us to 
perform  inversion for spin-one projectiles  where there is no
coupling between $l$ channels for given $j$  and 
we can justify the  omission of a tensor interaction.
We can  incorporate {\em a priori\/}
information, allowing us to invert when the amount of
information to fix each potential component is quite low (e.g.\ the use of
the 37.33 MeV solution in the 19.6 MeV case).

{\bf The potentials.}\,\,
We have determined potentials representing the KKST $S_{lj}$ for 37.33 and 19.6
MeV. The real components are remarkably similar apart from 
the expected energy dependence. General features are consistent
with those derived from a spinless model, and there is every reason
to believe the potentials are reliable.

It is interesting therefore that the various imaginary components at the different
energies differ quite markedly, particularly for
$r<1$. Although  the potential components are least
reliably determined for $r<1$ fm, and do possibly interact with each other in
the inversion process, it is a firm conclusion  that the imaginary potential 
is of very different form at 19.6 and 37.3 MeV. This includes notably  the large 
magnitude of the central $V_1$ term at smaller $r$. The larger number of open 
channels at the higher energies makes such a difference in imaginary
potential  plausible. However  the form of the imaginary potential
must certainly be influenced by the fact that
only $\Delta l =0$ transfer was included  in the KKST  calculations. 
CDCC calculations~\cite{cdcc} at higher energies
suggest~\cite{aairsm}  that breakup  into $l=2$ states of the 
deuteron modifies the absorptive potentials at larger radii.

In view of the uncertainty associated with fitting $S_{lj}$ for spin-one projectiles
with a vector spin-orbit potential alone, it is interesting that a universal property
of the real spin-orbit $V_1$ term has emerged. This term is, in every case, 
whether there is deuteron 
distortion or not, essentially the same: i.e.\ a surprisingly deep potential hole
at $r=0$. The corresponding $V_2$ term nearly always shares this property. 
The interpretation of this requires further study;
perhaps it is the consequence of representing with a vector potential
that which would be better represented by a tensor potential. The fact that
in the present case
$S_{lj}$ can indeed be represented by a vector term alone points to the ambiguity 
problems which lie ahead for studies involving tensor potentials.  

{\bf Conclusions relating to RGM and MCRGM.}\,\,
The results here are no more than a first step in 
extracting information concerning the deuteron-nucleus potential 
from RGM calculations. The possibilities for
studying the way different NN potentials or different reaction channels
included in the RGM model manifest themselves in the single particle
local potential are obvious. The model of KKST leads to a
potential with certain broad features corresponding to the phenomenological
Woods-Saxon based OM, but having differences in detail. 
In particular, the RMS radius of the real component
is larger than the OM value, although not too much weight should be
given to Woods-Saxon phenomenology with mediocre fits. It is satisfying that the volume
integral of the KKST imaginary central potential rises with 
increasing energy as the volume integral of the real part falls. It is also
appropriate that the latter falls rather more slowly than 
that for the nucleon potential. The contrary behaviour {\em in the surface, only\/},
of the  potential derived from Lemere {\em et al\/} $S_{lj}$ is one of many
observations deserving systematic study.  The need for MCRGM calculations 
which include higher $l$-transfer to $l>0$ deuteron virtual states is a high priority 
if it is true that the CDCC findings apply in general.

{\bf Conclusions relating to deuteron-nucleus interactions.}\,\,
Caution is needed in extrapolating from the deuteron-$\alpha$ interaction
to general properties of deuteron-nucleus potentials. Indeed, the magnitude
of the
parity dependent components will fall with increasing  $A$ of the target 
nucleus. Nevertheless, the present study does 
suggest that deuteron-nucleus potentials  have certain features which
are not generally allowed for in phenomenological studies. One such feature
is the very deep dip near $r=0$ in the imaginary potential, possibly
related to unusual non-local or $l$-dependence effects. 
The lowest order deuteron folding model implies~\cite{kunz}
that the non-locality range of the deuteron-nucleus potential is
half that of the folded-in nucleon-nucleus non-locality. 
But the present work shows that a fully antisymmetrized system with channel coupling
involves other forms of non-locality as well. The  `anti-Perey' effect,
possibly due to two-nucleon exchange terms in the kernel, is evidence for this.

{\bf General conclusions concerning inter-nucleus potentials.}\,\,
The results presented here suggest generic properties which are almost
never represented in conventional phenomenology based on Woods-Saxon potentials.
While some of these properties will certainly be attenuated for heavier
target nuclei, they show that conventional phenomenology does not 
provide a sound basis for the comparative evaluation of relativistic
and non-relativistic approaches.  

{\bf Future RGM calculations.}\,\,
A systematic RGM study combined with inversion could clearly yield
much information concerning both RGM and inter-nuclear interactions.
Some steps in this direction are underway~\cite{shirley}, but the 
questions raised by the present work require very substantial RGM
calculations indeed. In the first place, the clear indication of
CDCC calculations that the $l=2$ states of the virtual deuteron are
important imply that these should also be included in RGM. Moreover,
CRC calculations of deuteron scattering on heavier nuclei~\cite{atall}
suggest that coupling to mass the three reaction channels play an important
role. This unexpected result requires confirmation in a fully
antisymmetrized RGM calculation in which the various doubts concerning
non-orthogonal channels which plague CRC calculations do not enter. 
Finally, there remains the  challenge of studying the consequences of using
more realistic potentials (tensor interaction, repulsive cores, etc.),
see for example Ref.~\cite{hofmann}.

\section*{Acknowledgements}
We are most grateful Professor Y. C. Tang for tabulated phase shifts and
helpful advice and  to the Engineering and Physical Science Research Council
of the UK for grants.

\newpage
{\sc Figure Captions}

\setlength{\parindent}{0.0 in}
\setlength{\parskip}{0.3 in}

{\sc Figure 1.} Potentials found by inverting KKST $S_{lj}$ for
deuterons scattering from $^4$He. The dashed line represents
the pot3 solution for 37.33 MeV CM, and the solid line
solution pot-a for 19.6 MeV. These are the $V_1$, i.e.
parity independent components
arranged in order from from the top:  real central, imaginary central,
real spin-orbit and imaginary spin-orbit.
 
{\sc Figure 2.} The parity dependent, $V_2$, components for the same
cases as shown in Figure 1.

{\sc Figure 3.} Comparison of $V_1$ components for the pot3 (solid lines) and
potG1 (dashed lines) solutions for 37.33 MeV deuterons.

{\sc Figure 4.} Comparison of $V_2$ components for the pot3 (solid lines) and
potG1 (dashed lines) solutions for 37.33 MeV deuterons.

{\sc Figure 5.} The parity independent, $V_1$, components for 17 MeV (solid), 
25 MeV (dashed)  and 54 MeV (dotted)
deuterons scattering from $^4$He, derived by inverting the $S_{lj}$ of
Lemere {\em et al\/}.

{\sc Figure 6.} As for Figure 5, but here the $V_2$ components are plotted.


\begin{thebibliography}{99}
\bibitem{invprob}S.G. Cooper and R.S. Mackintosh, Inverse Problems, {\bf 5}
(1989) 707
\bibitem{prc54}S.G. Cooper and R.S. Mackintosh, Phys. Rev. {\bf C54} (1996) 
3133
\bibitem{shirley}S.G. Cooper, `Antisymmetrization and channel coupling
effects on the imaginary potentials for p $+\alpha$/d $+^3$He' OU preprint OUPD-9707
\bibitem{kkst}H. Kanada, T. Kaneko, S.Saito and Y.C. Tang, Nucl. Phys. 
{\bf A444} (1985) 209
\bibitem{ka}P.W. Keating and D.D. Armstrong, Phys. Rev. {\bf C8} (1973) 1692
\bibitem{grsnp21}G.R. Satchler, Nucl. Phys. {\bf 21} (1960) 116
\bibitem{bar}B.A. Robson, The Theory of Polarization Phenomena, (Clarendon
Press, Oxford, 1974)
\bibitem{andy}A.A. Ioannides and R.C. Johnson, Phys. Rev. {\bf C17} (1978)
1331
\bibitem{stamp}A.P. Stamp, Nucl. Phys {\bf A159} (1970) 399
\bibitem{chadsab}K. Chadan and P.C. Sabatier, Inverse Problems in Quantum 
Scattering Theory, Second Edition, (Springer, New York, 1989)
\bibitem{badhon}R.S. Mackintosh and S.G. Cooper, Invited Contrib. to Int. 
Symp. on Quantum Inversion Theory and
Applications, Bad Honnef {\em Springer Verlag, Berlin, 1994\/}
\bibitem{yct-priv} Y. C. Tang, private communication.
\bibitem{npa320}M. LeMere, D.J. Studeba, H. Horiuchi and Y.C. Tang,
  Nucl. Phys. {\bf A320} (1979) 449
\bibitem{tt-prc8}D.R. Thompson and Y.C. Tang, Phys. Rev. {\bf C8} (1973) 1649
\bibitem{mc}R.S. Mackintosh and S.G. Cooper J. Phys. G. {\bf 23} (1997) 565
\bibitem{pb}F. Perey and B. Buck, Nucl. Phys. {\bf 32} (1962) 353
\bibitem{perey}F.G. Perey, in: Direct interactions and nuclear reaction mechanisms, 
eds. E. Clementel and C. Villi (Gordon and Breach, New York, 1963);
N. Austern, Phys. Rev. {\bf B137} (1965) 752
\bibitem{kuku}V.I. Kukulin, private communication.
\bibitem{hinten}F. Hintenberger, G. Mairle, U. Schmidt-Rohr, G.J. Wagner
and P. Tureck,  Nucl. Phys. {\bf A111} (1968) 265
\bibitem{cdcc}M. Yahiro, M. Nakano, Y.Ieri and M. Kamimura,
Prog. Theor. Phys. {\bf 67} (1982) 1038; H. Amakawa, A. Mori, H. Nishioka, K. Yazaki
and S. Yamaji, Phys. Rev. {\bf C 23} (1981) 583; I.J. Thompson and M. Nagarajan,
Phys. Lett. {\bf 106B} (1981) 163
\bibitem{aairsm}A.A. Ioannides and R.S. Mackintosh, Phys. Lett.
{\bf 169B} (1986) 113
\bibitem{kunz}P.D. Kunz, Phys. Lett. {\bf 35B} (1971) 16
\bibitem{ltt}M. Lemere, Y.C. Tang and D.R. Thompson, Nuclear Physics
{\bf A266} (1976) 1

\bibitem{atall}R.S. Mackintosh, S.G. Cooper and A.A. Ioannides,
Nucl. Phys. {\bf A472} (1987) 85
\bibitem{hofmann}H.M. Hofmann and G.M.Hale, Nucl. Phys. {\bf A617}
(1997) 69

\end{thebibliography}
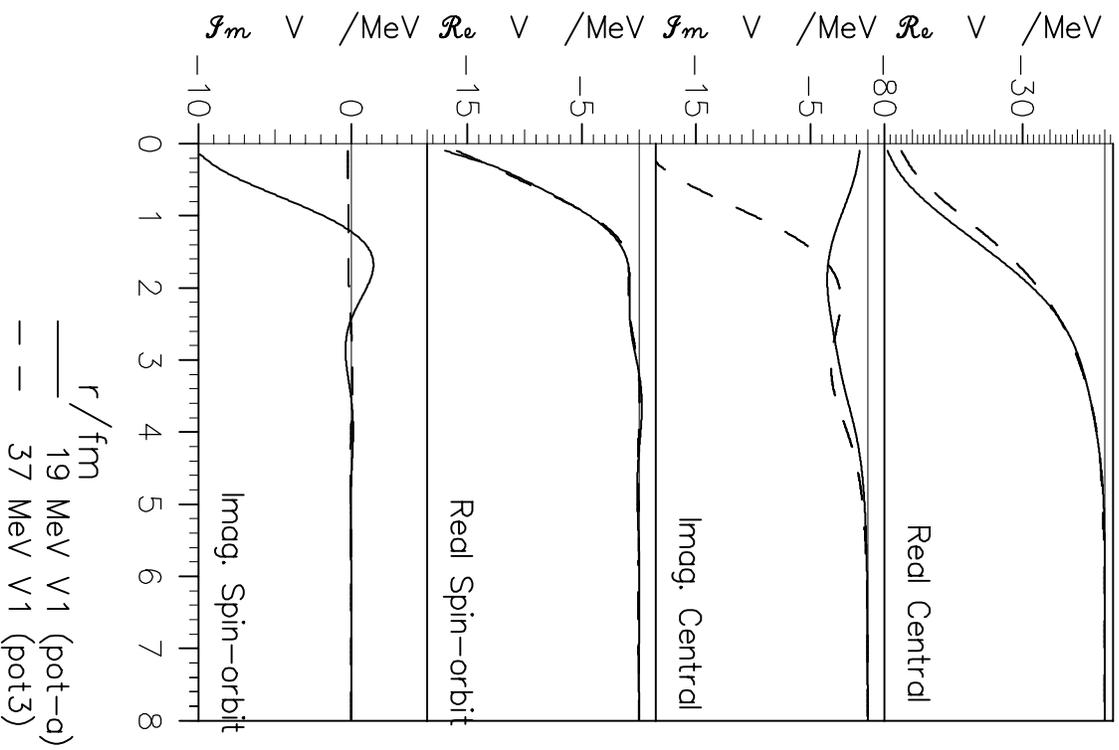
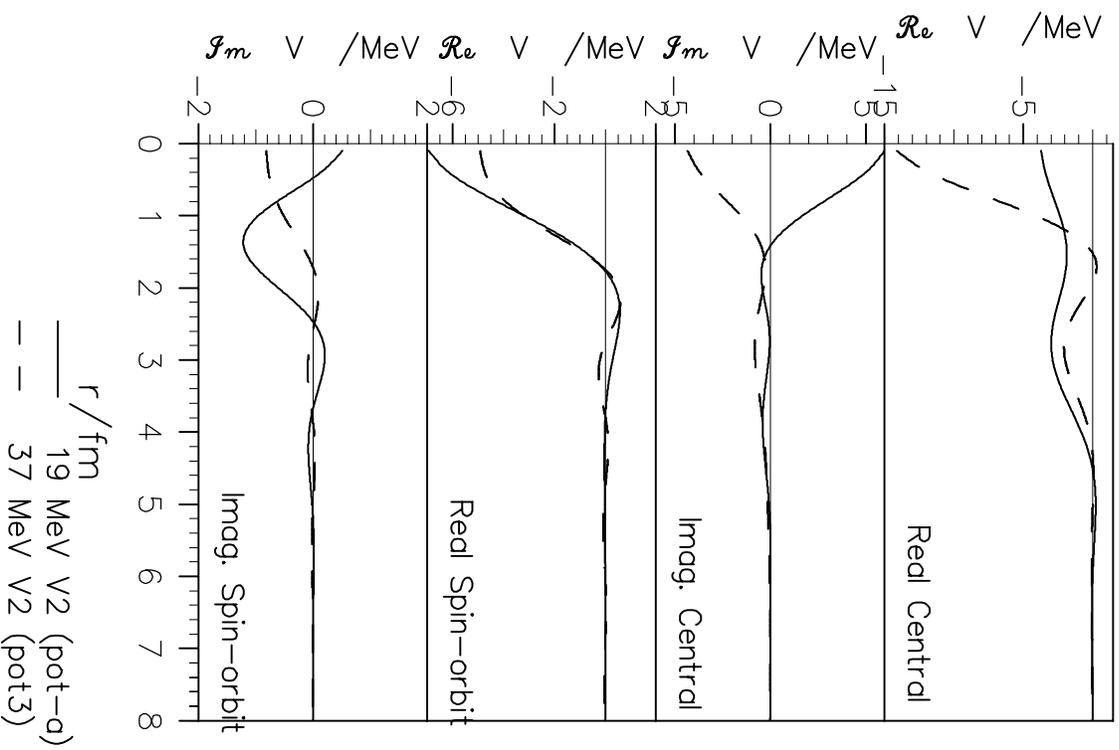
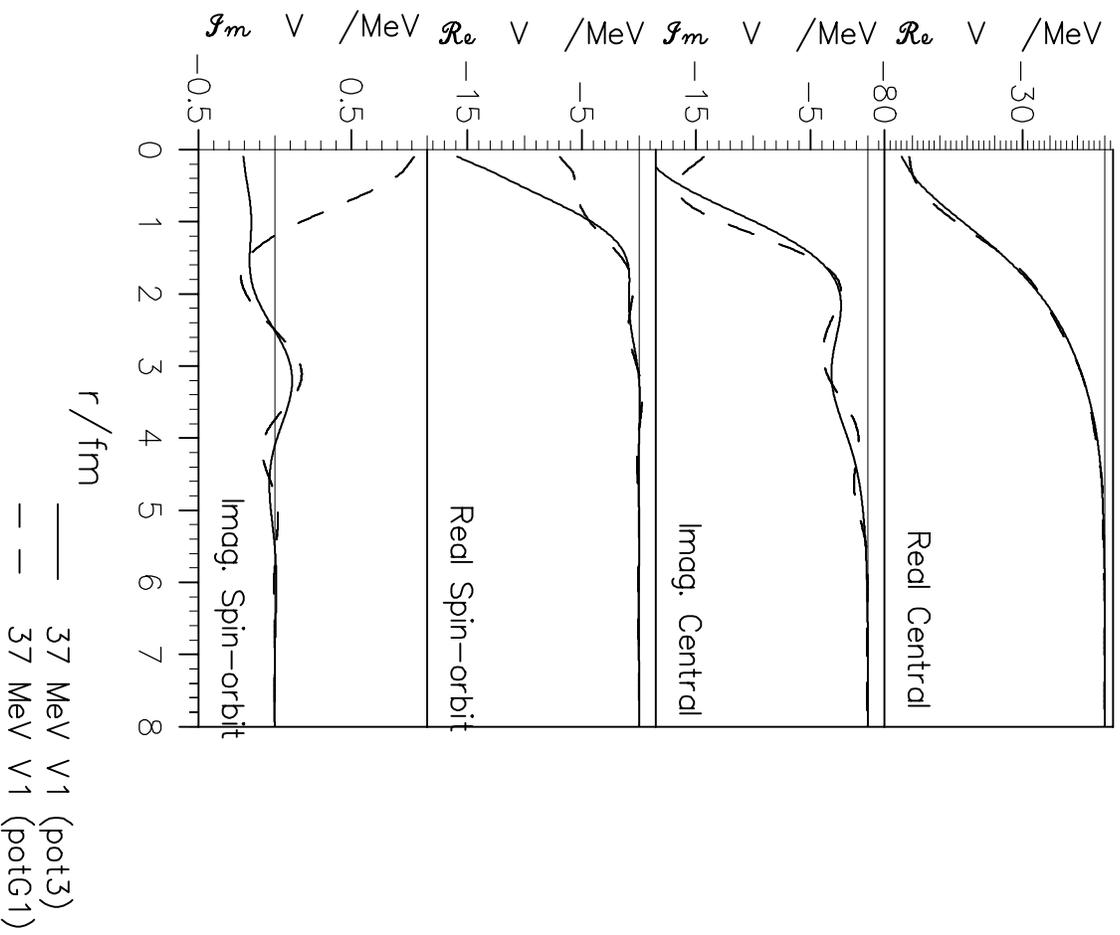
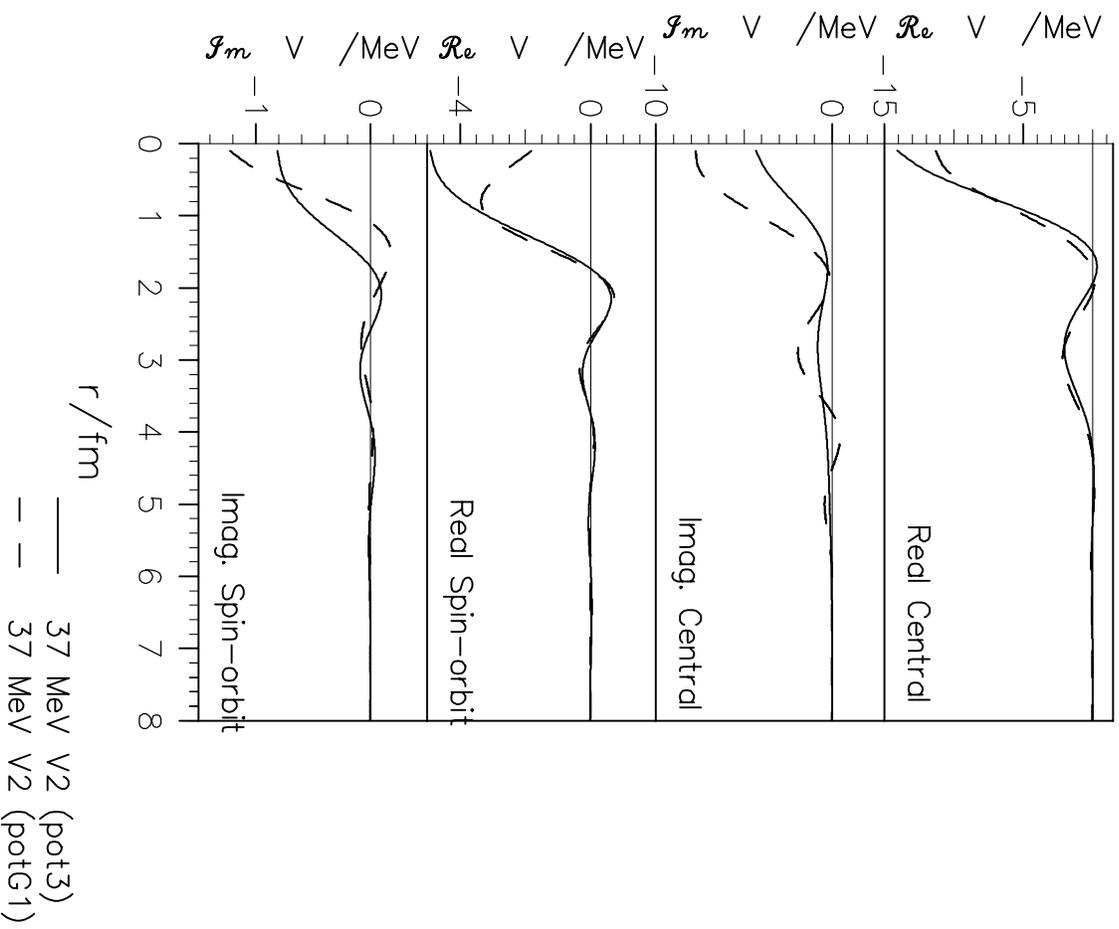
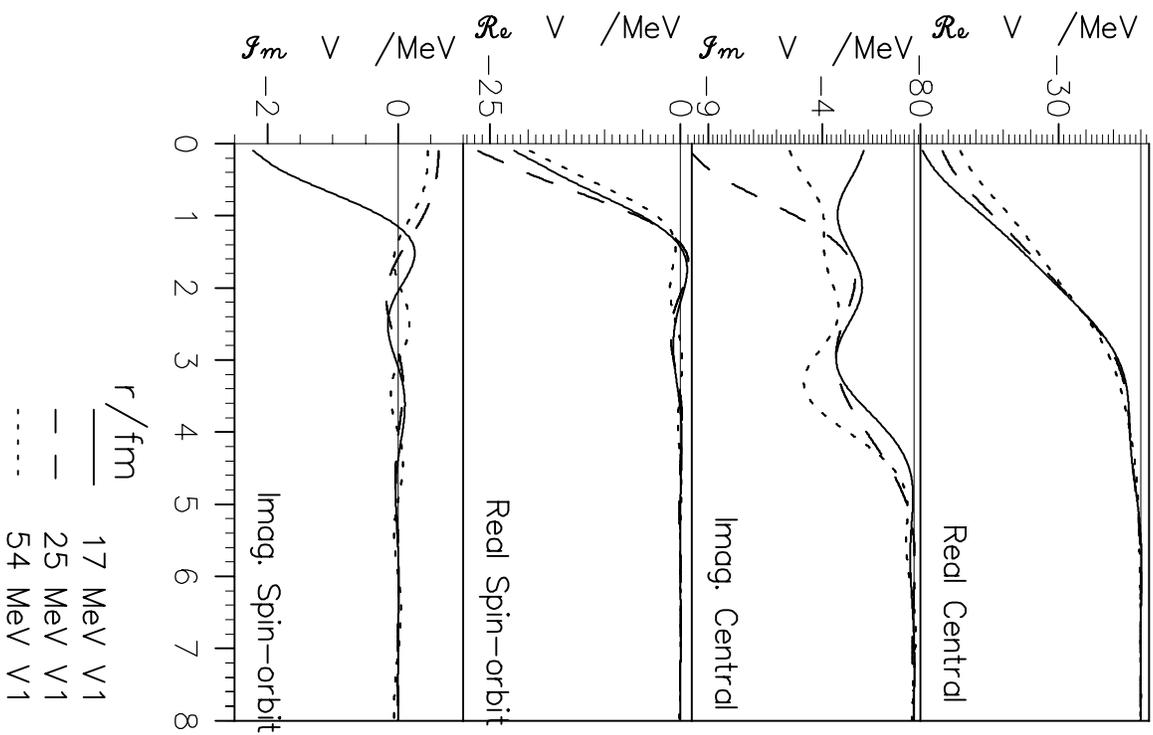
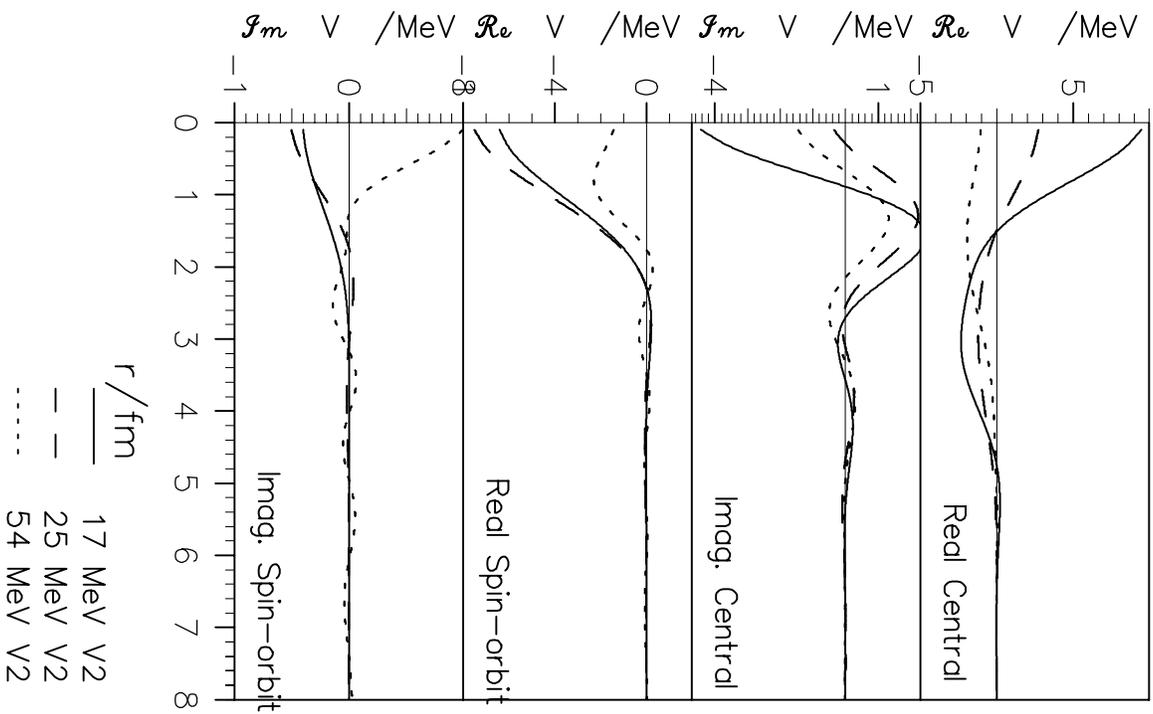
\end{document}